\documentclass[preprint, superscriptaddress, notitlepage]{revtex4-2}
\usepackage{amsfonts}
\usepackage{amssymb}
\usepackage[english]{babel}
\usepackage{hyperref}   
\usepackage{color}
\usepackage{float}
\usepackage{placeins}
\usepackage{amsmath}
\usepackage{hyphenat}
\usepackage{graphicx}  
\usepackage{caption}
\usepackage{subcaption}
\usepackage{siunitx}
\usepackage[symbol]{footmisc}
\usepackage{natbib}

\newcommand{\bn}{\mbox{\boldmath $n$}}
\newcommand{\bN}{\mbox{\boldmath $N$}}

\newcommand{\bx}{\mbox{\boldmath $x$}}
\newcommand{\bu}{\mbox{\boldmath $u$}}

\newcommand{\bB}{\mbox{\boldmath $B$}}
\newcommand{\br}{\mbox{\boldmath $r$}}

\newcommand{\bj}{\mbox{\boldmath $j$}}
\newcommand{\be}{\mbox{\boldmath $e$}}
\newcommand{\bv}{\mbox{\boldmath $v$}}

\newcommand{\bG}{\mbox{\boldmath $G$}}
\newcommand{\bD}{\mbox{\boldmath $D$}}

\begin{document}
\raggedbottom 
\title{Field configurable skyrmion devices utilising fine planar motion control}

\author{Charles Kind}
\affiliation{Department of Computer Science, University of Bristol, Bristol BS8 1UB, UK}

\date{\today}

\begin{abstract}
Magnetic skyrmions are nano-scale spin textures whose stability, size and ease of manipulation make them strong contenders for next generation data and logic applications. Here we illustrate fine motion control of skyrmions, we show how they can be moved in any direction and for any distance, in the plane. We demonstrate, using simulations, two novel skyrmionic devices whose function is predicated upon the fine planar motion control of skyrmions. The first device is a field configurable logic gate that could be used as a component in a larger skyrmionic processor. The second is a field configurable skyrmion spiking neuron device that could be a component in larger skyrmionic neural networks. Such devices are highly energy efficient and are potentially useful in the field of neuromorphic computing, which seeks to build 'brainlike' computers. 
Finally we explore the concept that this free planar motion, which enables us to ignore the skyrmion Hall effect, offers an important new paradigm in skyrmionic device theory moving away from the current applications such as racetrack memory.

\end{abstract}

\maketitle
\section{Introduction}

Magnetic skyrmions are topologically stable, particle-like, spin textures in the magnetization of chiral magnets \citep{bogdanovThermodynamicallyStableMagnetic1994}. They have been proposed as a possible route to a new generation of low power, high density memory \cite{fertSkyrmionsTrack2013, zhangMagneticSkyrmionLogic2015} and as candidates for use in neuromorphic computing systems \cite{chenSkyrmionicInterconnectDevice2020, songSkyrmionbasedArtificialSynapses2020,grollierNeuromorphicSpintronics2020}.

Many different types of skyrmionic device have been considered since the idea of using them as media for data and logic applications was first introduced. The predominant type of device considered is based upon the concept of magnetic racetrack first introduced by Stuart Parkin et al. in 2008 \cite{parkinMagneticDomainWallRacetrack2008} and put a skyrmionic perspective by Albert Fert et al. in 2013 \cite{fertSkyrmionsTrack2013}. The basic idea is that we push skyrmions down a thin wire exploiting the low electrical currents required to drive them to maximise their velocities. Scientists have considered ferrimagnetic \cite{wooCurrentdrivenDynamicsInhibition2018} and antiferromagnetic \cite{zhangAntiferromagneticSkyrmionStability2016} domains to overcome potentially undesirable physical features of skyrmions such as the topological Hall effect \cite{jiangDirectObservationSkyrmion2017} which tends to push the skyrmion at an angle from the direction of electron flow. More recently skyrmions have been considered in the context of logic-in-memory architecture \cite{gnoliSkyrmionLogicInMemoryArchitecture2021} and as artificial neurons \cite{chenSkyrmionicInterconnectDevice2020}, though still using one dimensional (1D) current flow to drive motion. In this work we consider full two dimensional (2D) skyrmion motion in the plane. We demonstrate using analysis of the Thiele \cite{thieleSteadyStateMotionMagnetic1973} equation and using simulations that fine 2D motion control of skyrmions is possible and can effectively ignore the Hall effect even in ferromagnetic systems. Furthermore, we model two simple devices that exploit this mechanism demonstrating the impact this shift in perspective will have on emerging spintronics applications.

%
%

Magnetic skyrmions can be found in the magnetization of certain chiral magnets \citep{bogdanovThermodynamicallyStableMagnetic1994}. Present in both bulk magnetic materials and interfacially in magnetic multilayers, they are stabilised by the Dzyaloshinkii–Moriya interaction (DMI) \cite{dzyaloshinskyThermodynamicTheoryWeak1958, moriyaAnisotropicSuperexchangeInteraction1960}. The magnetisation vectors of a single skyrmion, in the two dimensional (2D) continuous model, are a cover of the two-sphere and carry a unitary degree, $|Q|=1$, defined as
\begin{align} \label{TopDeg}
Q = \frac{1}{4 \pi} \int \bn \cdot \left( \partial_1 \bn \times \partial_2 \bn \right) \: d^2 x, 
\end{align}
where $\bn \left( \bx \right)$ is the unit vector field of magnetization. Under this definition, a single stable skyrmion has the degree $Q=-1$. This quantized degree is the reason for the topological stability of the skyrmion. In the continuous mathematical model the energy cost for moving away from this state is infinite but in real magnetic systems the spins are defined on a discrete lattice at the atomic level and samples are finite in extent meaning that the energy barrier is also finite. Nonetheless the energy barrier is real and high enough that skyrmions in the systems we consider can be thought of as stable with respect to perturbing energies below that barrier.

\begin{figure*}[]
\centering
\includegraphics[width = .95\columnwidth]{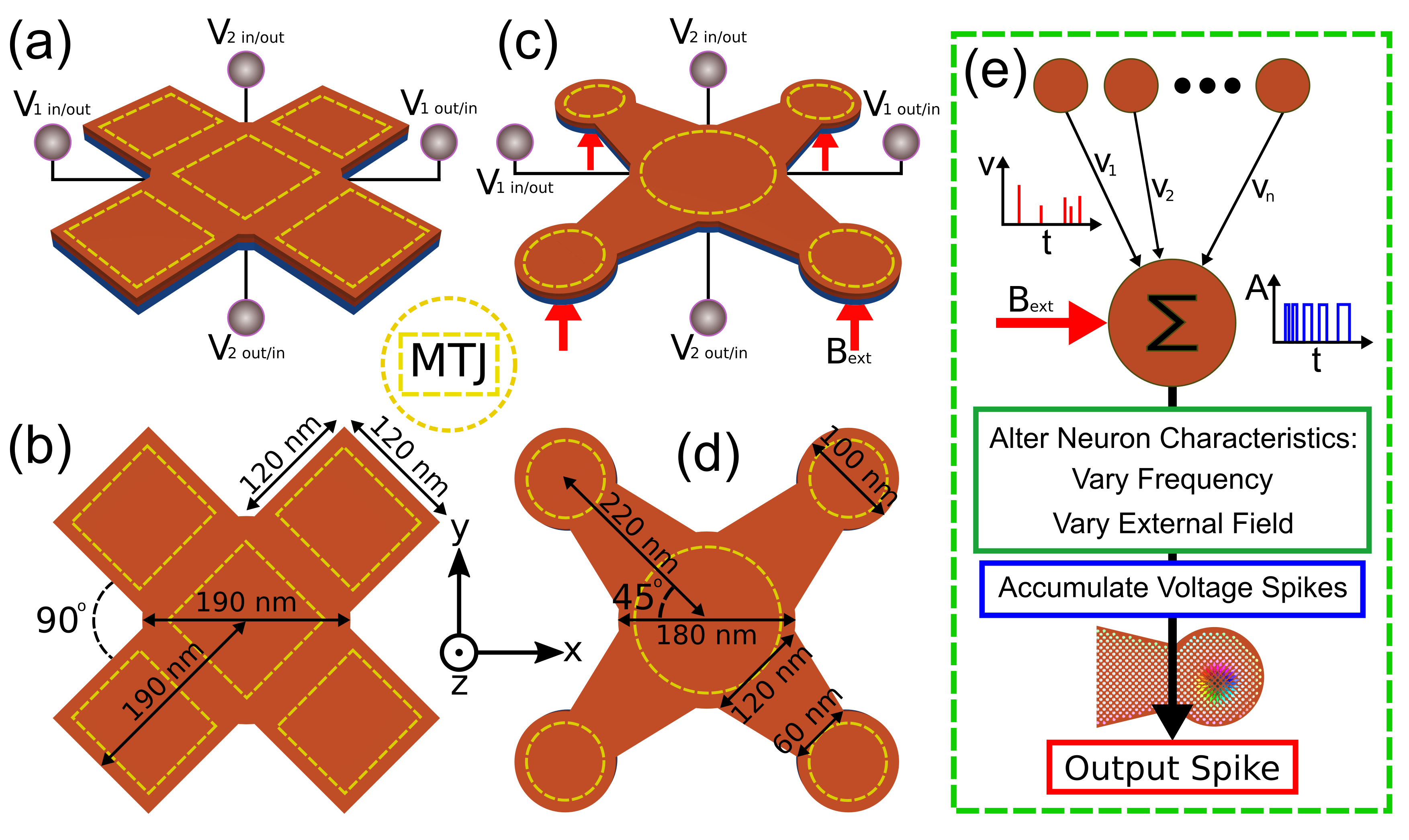}
\caption{\textbf{A skyrmion configurable logic gate and configurable neural network node.}
   \textit{	 
   \textbf{(a)} A sketch of a skyrmion configurable logic gate (SCLG). The dashed yellow areas represent magnetic tunnel junctions where changes in topological degree cause quantized changes in resistivity thus measuring skyrmion location. Note the two pairs of electrical contacts, each can pair can swap the source and sink.
   \textbf{(b)} A top view of the junction from (a) with planar dimensions included. 
   \textbf{(c)} A sketch of a skyrmion configurable neural network (SCNN) node. The dashed yellow areas represent magnetic tunnel junctions. There are two pairs of electrical contacts as in (a). Here we also see four red arrows representing separate localised external magnetic fields ($B_{ext}$) applied to each arm of the device. 
   \textbf{(d)} A top view of the junction from (c) with planar dimensions included.
   \textbf{(e)} A schematic of a network of SCNNs. The node Sigma $\left( \sum \right)$ accumulates voltage inputs from presynaptic nodes and fires at a threshold. The external field and local frequency of the node can be altered to change its firing characteristics. The node 'fires' when the skyrmion enters the end bulb.
  }}
\label{fig1}
\end{figure*}

Free planar motion is achieved through the pulsing of current at high frequency in directions orthogonal to each other. This then forms a basis and the almost linear response of skyrmion velocity to applied current is exploited to derive the required vectors. The setup of the two devices is shown in Fig. \ref{fig1}. The first device, Fig. \ref{fig1} (a), is a skyrmion configurable logic gate (SCLG) and the second, Fig. \ref{fig1} (c), a skyrmion configurable neural network (SCNN) node based on the work of Xing Chen et al. \cite{chenCompactSkyrmionicLeaky2018}. Detailed explanations of their form and function will follow after we explore the theory of 2D motion.


\section{Skyrmion motion theory}

We apply the Thiele \cite{thieleSteadyStateMotionMagnetic1973} approach to the dynamics of our skyrmionic magnetisation textures assuming the stationary limit, where the magnetisation texture moves with constant velocity, and assuming that the texture does not deform. We consider the current in plane (CIP) geometry using the Zhang-Li model \cite{zhangRolesNonequilibriumConduction2004}. The travelling wave ansatz, $\bn \left( \br , t \right) = \bn \left( \br - \bv_d t\right)$ , is then applied to the Landau-Lifshitz-Gilbert \cite{gilbertphenomenologicaltheorydamping2004} dynamical equation 
with torque to produce \cite{thiavilleMicromagneticunderstandingcurrentdriven2005a, everschorCurrentinducedRotationalTorques2011, eliasSteadymotionskyrmions2017}
Thiele's equation \cite{thieleSteadyStateMotionMagnetic1973}. 
\begin{align} \label{ThieleCIP}
\bG \times  \left( \bj - \bv \right) + \bD \left( \beta \bj - \alpha \bv \right) = 0,
\end{align}
here $\bG = 4 \pi Q \be_z$ is the gyrocoupling vector, 
$\bD$ is the dissipative tensor with components 
\begin{align} \label{DTensor}
D_{ij} = \int \partial_i \bn \cdot \partial_j \bn \: d^2x, 
\end{align}
$\bv$ is the velocity of the centre of mass of the skyrmion, $\bj$ is the current vector, $\alpha$ is the Gilbert damping constant and $\beta$ is the dissipative spin transfer torque parameter. For an axially symmetric texture the dissipative tensor $\bD$ will have components $D_{1,1} = D_{2,2}$ and $D_{i,j} = 0$, for $i \ne j$  \cite{HuberDynamicsspinvortices1982}. 

Solving Thieles equation, given $\bv = \left( \bv_x , \bv_y \right)$, we get
\begin{align} \label{TSolve}
\bv_x = \frac{\left( \alpha \beta D^2 + G^2\right) \bj_x - GD \left( \alpha - \beta \right) \bj_y}{\left( \alpha D \right)^2 + G^2}, \\
\bv_y = \frac{GD \left( \alpha - \beta \right) \bj_x + \left( \alpha \beta D^2 + G^2\right) \bj_y}{\left( \alpha D \right)^2 + G^2}.
\end{align}

If we now solve for two orthogonal currents, $\bj_1 = \left( \bj , 0 \right)$ and  $\bj_2 = \left( 0 , \bj \right)$, it is clear that the product of the resulting velocity vectors is zero.

We know that above the de-pinning threshold the current velocity relationship of skyrmions is close to linear \cite{iwasakiUniversalCurrentvelocityRelation2013, tomaselloStrategyDesignSkyrmion2014} and that if we reverse the the current flow we reverse the direction of motion \cite{wangElectricalManipulationSkyrmions2022}. We therefore measure the skyrmion velocity for two different current magnitudes, here $3 \times 10^{11}$A$\:$m$^{-2}$ and $6 \times 10^{11}$A$\:$m$^{-2}$, in both the $e_x$ and $e_y$ directions and derive unit vectors and average velocity per $1 \times 10^{10}$A$\:$m$^{-2}$. From these values, assuming linearity at the current levels applied, we can calculate a current magnitude and time required in each direction of current application, $e_x$ and $e_y$. The currents are then applied divided by the time-steps taken, here we switch the current at 200 MHz \cite{wooCurrentdrivenDynamicsInhibition2018}, to arrive at a prescription for moving a skyrmion too a specific point in the plane. As the deflection caused by the skyrmion Hall effect is included in the derived vectors it can effectively be ignored.
This process will work for any topological degree, Eq. (\ref{TopDeg}), of skyrmion including skyrmion bags \cite{fosterTwodimensionalSkyrmionBags2019,kindMagneticSkyrmionBinning2021} and other higher degree skyrmions as the variation in Hall deflection makes no difference for this closed system.

As the currents are pulsed in two separate directions the maximum attainable velocity is approximately half of the velocity attainable with current pulsed in a single direction. This can be reduced if steps are taken to optimise the velocity by increasing pulse magnitude and decreasing pulse time for the shorter vector direction however we do not explore this here.

We use the graphics processing unit (GPU) enabled micromagnetic solver Mumax3 \cite{vansteenkisteDesignVerificationMuMax32014} to test our model by applying current in the cardinal and intercardinal directions, $ \frac{2 \pi}{n} | n \in \left\lbrace 1 ... 8 \right\rbrace$, and for distances in the range $\left[ 2,40\right]$ nm in steps of $2$ nm. Simulation parameters are given in the appendix. We find that the predicted directions vary by less than 2\% and the distances by less than 1\%.
%

%
%
%

\section{Skyrmion configurable devices}
To demonstrate the effectiveness of this paradigm we simulate two different basic devices that use 2D skyrmion motion. These devices are designed such that they could be built into digital or analogue architectures as the primary required inputs are current magnitude and pulse duration and the outputs are current from the activation of magnetic tunnel junction (MTJ) switches. MTJ's can detect quantised changes in resistivity due to the presence of skyrmions \cite{guangElectricalDetectionMagnetic2023} and are considered ideal for the electrical detection of skyrmions. The ability to be deployed in analogue or digital architectures is important for use in neuromorphic applications and in solely analogue for use in high radiation environments such as fusion reactors and in space. We call them configurable as both the SCLG and the SCNN could be configured in the field to fulfil a number of roles.


\begin{figure*}[]
\centering
\includegraphics[width = 0.45\columnwidth]{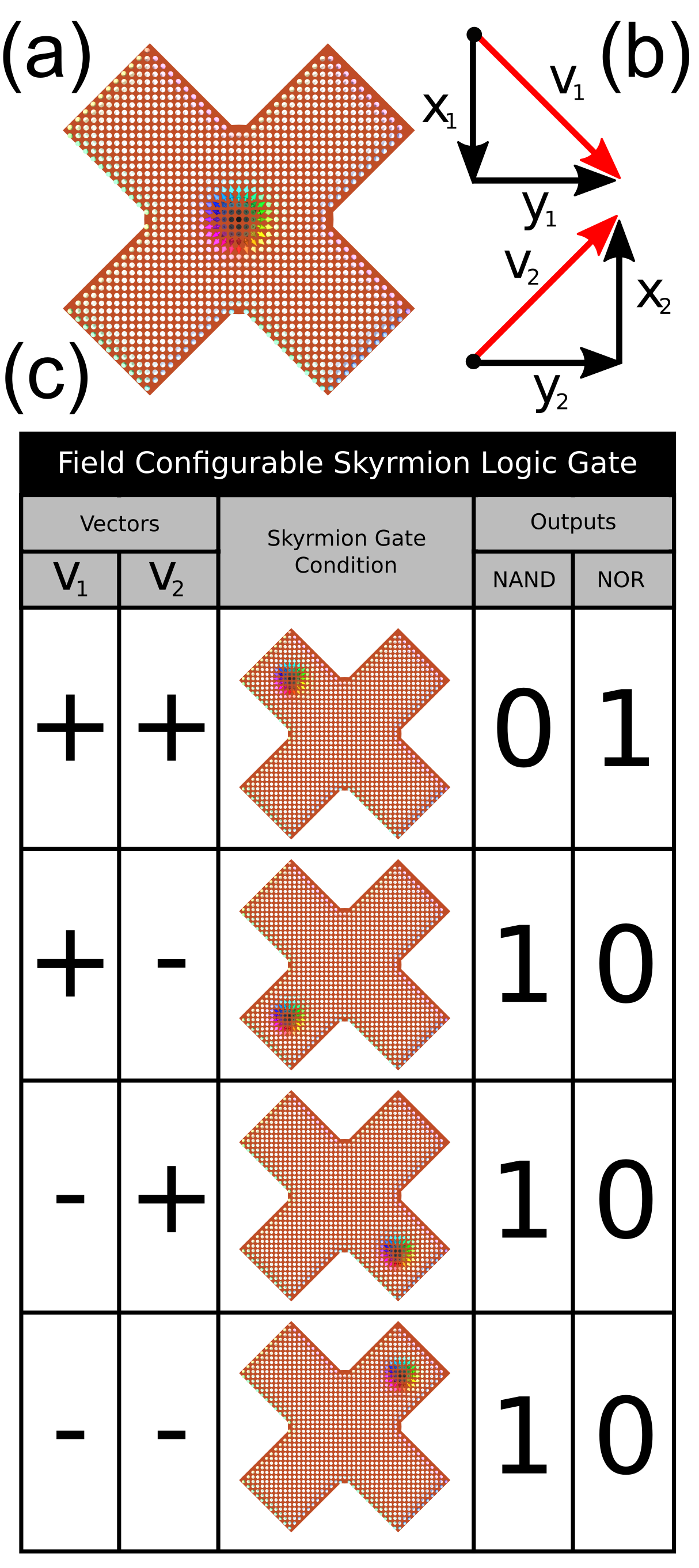}
\caption{\textbf{A field configurable skyrmion logic gate.}
   \textit{
   \textbf{(a)} A simulated logic gate with a skyrmion at its centre.
   \textbf{(b)} Direction vectors derived from the calculated basis vectors.
   \textbf{(c)} Logic table for current inputs and NAND and NOR gate outputs. NAND and NOR where chosen as they are the universal logic gates and can construct all others.
  }}
\label{fig2}
\end{figure*}


The SCLG can be preconfigured with a given logic gate configuration with each current pairing giving a different vector to a distinct MTJ, see Fig. \ref{fig2}, and each MTJ configured to provide a current response. We illustrate this in Fig. \ref{fig2} (c), providing examples of NAND and NOR, the universal logic gates. This is a very simple device, theoretically ideal for production en masse in a large scale field programmable processor.


\begin{figure*}[]
\centering
\includegraphics[width = 0.95\columnwidth]{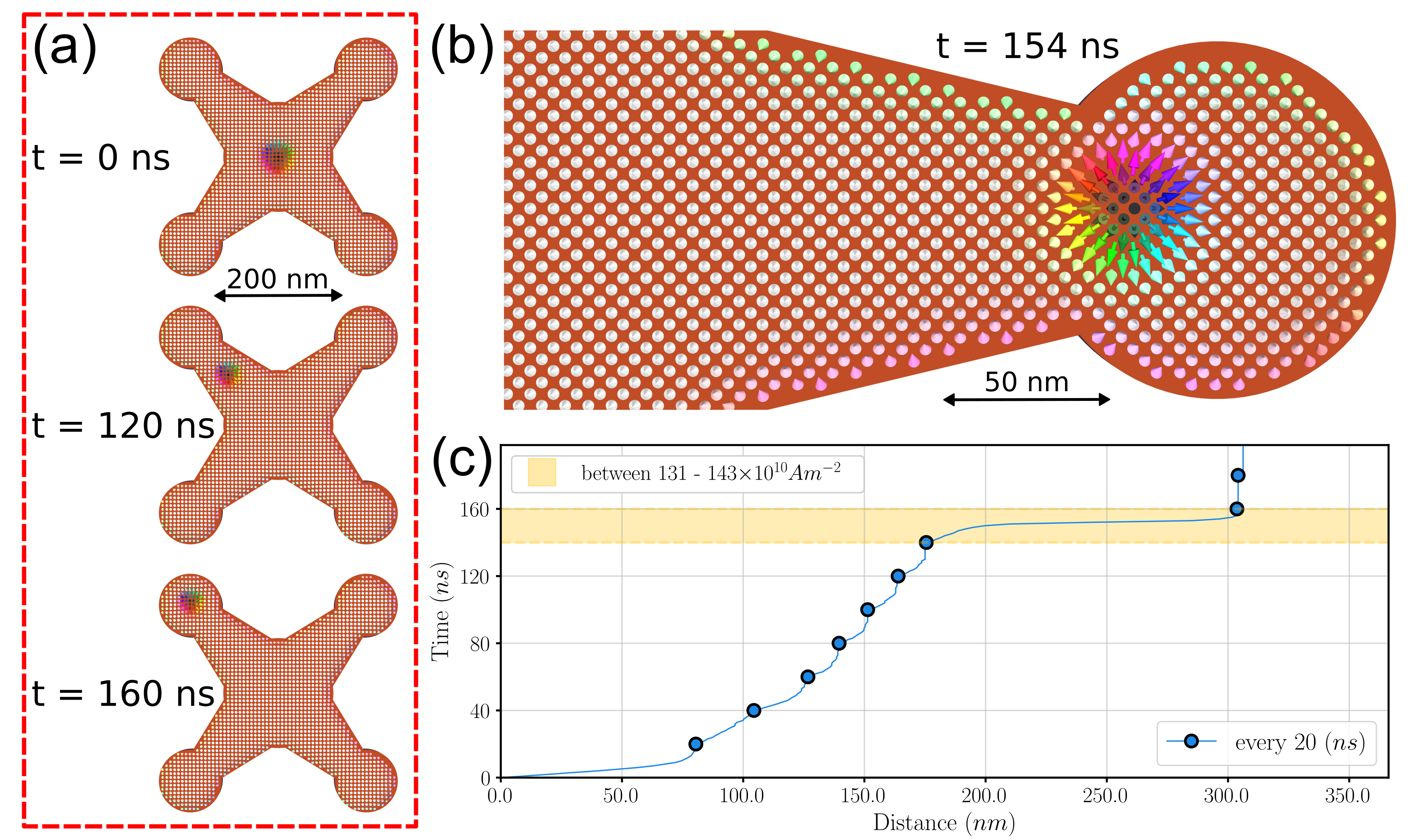}
\caption{\textbf{A field configurable skyrmion neural network node.} \textit{ Currents are accumulated from $6 \times 10^{11}$A$\:$m$^{-2}$ to MTJ activation at between $13.1$ and $14.3 \times 10^{11}$A$\:$m$^{-2}$ over a period of $160$ ns at zero external field.
   \textbf{(a)} The progress of the skyrmion to activation over the $160$ ns time period. 
   \textbf{(b)} A close up of the skyrmion entering the MTJ disc.
   \textbf{(c)} We plot the distance from the device center to the MTJ disc over time. Note that the skyrmion comes to approximate rest between later current increases prior to activation inside the yellow highlighted region.
  }}
\label{fig3}
\end{figure*}


The SCNN, see Fig. \ref{fig3}, is designed in a similar fashion to the SCLG and features four arms with MTJ detectors. This design requires additional external control in the form of an accumulator for the input currents, see Fig. \ref{fig1} (e). The skyrmion is driven by the accumulated currents from many pre-synaptic SCNN's until sufficient current is present to drive the skyrmion through the edge repulsion barrier that is the constriction at the end of the arm prior to the MTJ disc. This simulates the leaky–integrate-fire (LIF) spiking model of neurons \cite{abbottLapicqueIntroductionIntegrateandfire1999,hodgkinQuantitativeDescriptionMembrane1990} . Only one arm of the device can be utilised during any one activation sequence and each arm can provide a different activation profile by tuning the externally applied field, see Fig. \ref{fig1} (c).

Both devices feature MTJ's at their centres in order to detect the base state. In both cases the skyrmion can be driven, by an opposing applied pulse current, back to that base state. In the case of the SCNN this must be done to push the skyrmion from the MTJ activation site at the end of the arm as the constriction will not allow free passage. Once the skyrmion is free from the MTJ disc in the SCNN case and after activation in the SCLG case the skyrmion will return to the center of each device with no further applied current as this is the lowest energy state, this is however much slower than driving the skyrmion to the center. In both cases the returning current is lower than the activation current as the edge repulsion acts to drive the skyrmion toward the lowest energy state.

\section{Discussion}

We have shown that skyrmions can be driven with fine control in 2D. This extra dimension allows us to reconsider the design of many proposed skyrmionic devices. We use this 2D control to demonstrate skyrmion based configurable logic gates that could then be deployed in more complex information processing circuits. We also showed a configurable neural network node device that could have it's firing activation functions altered and could simulate four separate neurons at distinct times. These devices where chosen so that a complete next generation computing architecture could be considered, drawing from classical computing and neuron inspired brain-like computing.

One motivation behind this work was the authors attempt to drive skyrmions down arbitrarily shaped tree like structures that could be used to represent the dendritic arbors of a typical human neuron. In conclusion we hope that the shift from 1D to 2D skyrmionic motion will enable a new perspective in the application of skyrmions to modern and future data, logic and brain-like applications.

\section{Appendix}

We consider skyrmionic structures in thin film multilayers where the interfacial DMI results in N\'eel type skyrmions. The simulations where performed using the GPU-accelerated micromagnetic simulation program MuMax3 \cite{vansteenkisteDesignVerificationMuMax32014} with Landau-Lifshitz dynamics in the form
\begin{align}\label{llg}
\frac{\partial \bn}{\partial t}=\gamma\frac{1}{1+\alpha^2}\left(\bn \times \bB_{\mathrm{eff}}+\alpha\bn
\times\left( \bn \times \bB_{\mathrm{eff}} \right) \right),
\end{align}
where $\gamma \approx 176$ rad(ns$\:$T)$^{-1}$ is the electron gyromagnetic ratio, $\alpha=0.3$ the dimensionless damping parameter, $\bB_{\mathrm{eff}} = \delta E / \delta \bn$ the effective field and $\bn(\bx)=\bN (\bx)/M_{s}$ the magnetisation vector field normalised by the saturation magnetisation. 

The simulations external geometry is a $512\times512$ nm$^2$ rectangle of $1$ nm thickness, in order to represent a typical surface that could be fabricated using lithographic processing. Cell size of $2 \times 2 \times 1$ nm$^3$ has been used. Material parameters are: saturation magnetisation $M_{s} = 580$ kAm$^{-1}$, exchange $20$ pJm$^{-1}$, interfacial DMI $3$ mJm$^{-2}$ and uniaxial anisotropy along the $+z$ direction $0.53$ MJm$^{-3}$ \cite{kindExistenceStabilitySkyrmion2020}. 

The applied current is of Zhang-Li \cite{zhangRolesNonequilibriumConduction2004} (CIP) type, the non-adiabacity of spin transfer torque constant is set to $\xi = 0.05$, current flows in the $\be_x$ and $\be_y$ directions at separate times: 

\begin{align}\label{ZL}
\overrightarrow{\tau}_{ZL} &=\frac{1}{1+\alpha^2} \left( \left( 1 + \xi \alpha \right)\bn \times \left( \bn \times \left( \bu \cdot \nabla \right) \right) \bn  \nonumber \right. \\
&+ \left. \left( \xi - \alpha \right) \bn \times \left( \bu \cdot \nabla \right) \bn \right) \\
\bu &= \frac{\mu_B \mu_0}{2 e \gamma_0 B_{sat} \left( 1 + \xi^2 \right)}\bj
\end{align}

where $\bj$ A$\:$m$^{-2}$  is the current density, $\mu_B$ the Bohr magneton and $B_{sat}$ the saturation magnetisation expressed in Tesla.

\section*{Acknowledgements}
This work was supported by the Engineering and Physical Sciences Research Council (EPSRC) grant EP/M506473/1. 
The Titan V GPU used for parts of this research was donated by the NVIDIA Corporation.

\onecolumngrid

\bibliography{SkyrmDirect}

\begin{thebibliography}{31}%
\makeatletter
\providecommand \@ifxundefined [1]{%
 \@ifx{#1\undefined}
}%
\providecommand \@ifnum [1]{%
 \ifnum #1\expandafter \@firstoftwo
 \else \expandafter \@secondoftwo
 \fi
}%
\providecommand \@ifx [1]{%
 \ifx #1\expandafter \@firstoftwo
 \else \expandafter \@secondoftwo
 \fi
}%
\providecommand \natexlab [1]{#1}%
\providecommand \enquote  [1]{``#1''}%
\providecommand \bibnamefont  [1]{#1}%
\providecommand \bibfnamefont [1]{#1}%
\providecommand \citenamefont [1]{#1}%
\providecommand \href@noop [0]{\@secondoftwo}%
\providecommand \href [0]{\begingroup \@sanitize@url \@href}%
\providecommand \@href[1]{\@@startlink{#1}\@@href}%
\providecommand \@@href[1]{\endgroup#1\@@endlink}%
\providecommand \@sanitize@url [0]{\catcode `\\12\catcode `\$12\catcode
  `\&12\catcode `\#12\catcode `\^12\catcode `\_12\catcode `\%12\relax}%
\providecommand \@@startlink[1]{}%
\providecommand \@@endlink[0]{}%
\providecommand \url  [0]{\begingroup\@sanitize@url \@url }%
\providecommand \@url [1]{\endgroup\@href {#1}{\urlprefix }}%
\providecommand \urlprefix  [0]{URL }%
\providecommand \Eprint [0]{\href }%
\providecommand \doibase [0]{https://doi.org/}%
\providecommand \selectlanguage [0]{\@gobble}%
\providecommand \bibinfo  [0]{\@secondoftwo}%
\providecommand \bibfield  [0]{\@secondoftwo}%
\providecommand \translation [1]{[#1]}%
\providecommand \BibitemOpen [0]{}%
\providecommand \bibitemStop [0]{}%
\providecommand \bibitemNoStop [0]{.\EOS\space}%
\providecommand \EOS [0]{\spacefactor3000\relax}%
\providecommand \BibitemShut  [1]{\csname bibitem#1\endcsname}%
\let\auto@bib@innerbib\@empty
\bibitem [{\citenamefont {Bogdanov}\ and\ \citenamefont
  {Hubert}(1994)}]{bogdanovThermodynamicallyStableMagnetic1994}%
  \BibitemOpen
  \bibfield  {author} {\bibinfo {author} {\bibfnamefont {A.}~\bibnamefont
  {Bogdanov}}\ and\ \bibinfo {author} {\bibfnamefont {A.}~\bibnamefont
  {Hubert}},\ }\bibfield  {title} {\bibinfo {title} {Thermodynamically stable
  magnetic vortex states in magnetic crystals},\ }\href
  {https://doi.org/10.1016/0304-8853(94)90046-9} {\bibfield  {journal}
  {\bibinfo  {journal} {Journal of Magnetism and Magnetic Materials}\ }\textbf
  {\bibinfo {volume} {138}},\ \bibinfo {pages} {255} (\bibinfo {year}
  {1994})}\BibitemShut {NoStop}%
\bibitem [{\citenamefont {Fert}\ \emph {et~al.}(2013)\citenamefont {Fert},
  \citenamefont {Cros},\ and\ \citenamefont
  {Sampaio}}]{fertSkyrmionsTrack2013}%
  \BibitemOpen
  \bibfield  {author} {\bibinfo {author} {\bibfnamefont {A.}~\bibnamefont
  {Fert}}, \bibinfo {author} {\bibfnamefont {V.}~\bibnamefont {Cros}},\ and\
  \bibinfo {author} {\bibfnamefont {J.}~\bibnamefont {Sampaio}},\ }\bibfield
  {title} {\bibinfo {title} {Skyrmions on the track},\ }\href
  {https://doi.org/10.1038/nnano.2013.29} {\bibfield  {journal} {\bibinfo
  {journal} {Nature Nanotechnology}\ }\textbf {\bibinfo {volume} {8}},\
  \bibinfo {pages} {152} (\bibinfo {year} {2013})}\BibitemShut {NoStop}%
\bibitem [{\citenamefont {Zhang}\ \emph {et~al.}(2015)\citenamefont {Zhang},
  \citenamefont {Ezawa},\ and\ \citenamefont
  {Zhou}}]{zhangMagneticSkyrmionLogic2015}%
  \BibitemOpen
  \bibfield  {author} {\bibinfo {author} {\bibfnamefont {X.}~\bibnamefont
  {Zhang}}, \bibinfo {author} {\bibfnamefont {M.}~\bibnamefont {Ezawa}},\ and\
  \bibinfo {author} {\bibfnamefont {Y.}~\bibnamefont {Zhou}},\ }\bibfield
  {title} {\bibinfo {title} {Magnetic skyrmion logic gates: Conversion,
  duplication and merging of skyrmions},\ }\href
  {https://doi.org/10.1038/srep09400} {\bibfield  {journal} {\bibinfo
  {journal} {Scientific Reports}\ }\textbf {\bibinfo {volume} {5}},\ \bibinfo
  {pages} {9400} (\bibinfo {year} {2015})}\BibitemShut {NoStop}%
\bibitem [{\citenamefont {Chen}\ \emph {et~al.}(2020)\citenamefont {Chen},
  \citenamefont {Li}, \citenamefont {Pavlidis},\ and\ \citenamefont
  {Moutafis}}]{chenSkyrmionicInterconnectDevice2020}%
  \BibitemOpen
  \bibfield  {author} {\bibinfo {author} {\bibfnamefont {R.}~\bibnamefont
  {Chen}}, \bibinfo {author} {\bibfnamefont {Y.}~\bibnamefont {Li}}, \bibinfo
  {author} {\bibfnamefont {V.~F.}\ \bibnamefont {Pavlidis}},\ and\ \bibinfo
  {author} {\bibfnamefont {C.}~\bibnamefont {Moutafis}},\ }\bibfield  {title}
  {\bibinfo {title} {Skyrmionic interconnect device},\ }\href
  {https://doi.org/10.1103/PhysRevResearch.2.043312} {\bibfield  {journal}
  {\bibinfo  {journal} {Physical Review Research}\ }\textbf {\bibinfo {volume}
  {2}},\ \bibinfo {pages} {043312} (\bibinfo {year} {2020})}\BibitemShut
  {NoStop}%
\bibitem [{\citenamefont {Song}\ \emph {et~al.}(2020)\citenamefont {Song},
  \citenamefont {Jeong}, \citenamefont {Pan}, \citenamefont {Zhang},
  \citenamefont {Xia}, \citenamefont {Cha}, \citenamefont {Park}, \citenamefont
  {Kim}, \citenamefont {Finizio}, \citenamefont {Raabe}, \citenamefont {Chang},
  \citenamefont {Zhou}, \citenamefont {Zhao}, \citenamefont {Kang},
  \citenamefont {Ju},\ and\ \citenamefont
  {Woo}}]{songSkyrmionbasedArtificialSynapses2020}%
  \BibitemOpen
  \bibfield  {author} {\bibinfo {author} {\bibfnamefont {K.~M.}\ \bibnamefont
  {Song}}, \bibinfo {author} {\bibfnamefont {J.-S.}\ \bibnamefont {Jeong}},
  \bibinfo {author} {\bibfnamefont {B.}~\bibnamefont {Pan}}, \bibinfo {author}
  {\bibfnamefont {X.}~\bibnamefont {Zhang}}, \bibinfo {author} {\bibfnamefont
  {J.}~\bibnamefont {Xia}}, \bibinfo {author} {\bibfnamefont {S.}~\bibnamefont
  {Cha}}, \bibinfo {author} {\bibfnamefont {T.-E.}\ \bibnamefont {Park}},
  \bibinfo {author} {\bibfnamefont {K.}~\bibnamefont {Kim}}, \bibinfo {author}
  {\bibfnamefont {S.}~\bibnamefont {Finizio}}, \bibinfo {author} {\bibfnamefont
  {J.}~\bibnamefont {Raabe}}, \bibinfo {author} {\bibfnamefont
  {J.}~\bibnamefont {Chang}}, \bibinfo {author} {\bibfnamefont
  {Y.}~\bibnamefont {Zhou}}, \bibinfo {author} {\bibfnamefont {W.}~\bibnamefont
  {Zhao}}, \bibinfo {author} {\bibfnamefont {W.}~\bibnamefont {Kang}}, \bibinfo
  {author} {\bibfnamefont {H.}~\bibnamefont {Ju}},\ and\ \bibinfo {author}
  {\bibfnamefont {S.}~\bibnamefont {Woo}},\ }\bibfield  {title} {\bibinfo
  {title} {Skyrmion-based artificial synapses for neuromorphic computing},\
  }\href {https://doi.org/10.1038/s41928-020-0385-0} {\bibfield  {journal}
  {\bibinfo  {journal} {Nature Electronics}\ }\textbf {\bibinfo {volume} {3}},\
  \bibinfo {pages} {148} (\bibinfo {year} {2020})}\BibitemShut {NoStop}%
\bibitem [{\citenamefont {Grollier}\ \emph {et~al.}(2020)\citenamefont
  {Grollier}, \citenamefont {Querlioz}, \citenamefont {Camsari}, \citenamefont
  {{Everschor-Sitte}}, \citenamefont {Fukami},\ and\ \citenamefont
  {Stiles}}]{grollierNeuromorphicSpintronics2020}%
  \BibitemOpen
  \bibfield  {author} {\bibinfo {author} {\bibfnamefont {J.}~\bibnamefont
  {Grollier}}, \bibinfo {author} {\bibfnamefont {D.}~\bibnamefont {Querlioz}},
  \bibinfo {author} {\bibfnamefont {K.~Y.}\ \bibnamefont {Camsari}}, \bibinfo
  {author} {\bibfnamefont {K.}~\bibnamefont {{Everschor-Sitte}}}, \bibinfo
  {author} {\bibfnamefont {S.}~\bibnamefont {Fukami}},\ and\ \bibinfo {author}
  {\bibfnamefont {M.~D.}\ \bibnamefont {Stiles}},\ }\bibfield  {title}
  {\bibinfo {title} {Neuromorphic spintronics},\ }\href
  {https://doi.org/10.1038/s41928-019-0360-9} {\bibfield  {journal} {\bibinfo
  {journal} {Nature Electronics}\ }\textbf {\bibinfo {volume} {3}},\ \bibinfo
  {pages} {360} (\bibinfo {year} {2020})}\BibitemShut {NoStop}%
\bibitem [{\citenamefont {Parkin}\ \emph {et~al.}(2008)\citenamefont {Parkin},
  \citenamefont {Hayashi},\ and\ \citenamefont
  {Thomas}}]{parkinMagneticDomainWallRacetrack2008}%
  \BibitemOpen
  \bibfield  {author} {\bibinfo {author} {\bibfnamefont {S.~S.~P.}\
  \bibnamefont {Parkin}}, \bibinfo {author} {\bibfnamefont {M.}~\bibnamefont
  {Hayashi}},\ and\ \bibinfo {author} {\bibfnamefont {L.}~\bibnamefont
  {Thomas}},\ }\bibfield  {title} {\bibinfo {title} {Magnetic {{Domain-Wall
  Racetrack Memory}}},\ }\href {https://doi.org/10.1126/science.1145799}
  {\bibfield  {journal} {\bibinfo  {journal} {Science}\ }\textbf {\bibinfo
  {volume} {320}},\ \bibinfo {pages} {190} (\bibinfo {year}
  {2008})}\BibitemShut {NoStop}%
\bibitem [{\citenamefont {Woo}\ \emph {et~al.}(2018)\citenamefont {Woo},
  \citenamefont {Song}, \citenamefont {Zhang}, \citenamefont {Zhou},
  \citenamefont {Ezawa}, \citenamefont {Liu}, \citenamefont {Finizio},
  \citenamefont {Raabe}, \citenamefont {Lee}, \citenamefont {Kim},
  \citenamefont {Park}, \citenamefont {Kim}, \citenamefont {Kim}, \citenamefont
  {Lee}, \citenamefont {Lee}, \citenamefont {Choi}, \citenamefont {Min},
  \citenamefont {Koo},\ and\ \citenamefont
  {Chang}}]{wooCurrentdrivenDynamicsInhibition2018}%
  \BibitemOpen
  \bibfield  {author} {\bibinfo {author} {\bibfnamefont {S.}~\bibnamefont
  {Woo}}, \bibinfo {author} {\bibfnamefont {K.~M.}\ \bibnamefont {Song}},
  \bibinfo {author} {\bibfnamefont {X.}~\bibnamefont {Zhang}}, \bibinfo
  {author} {\bibfnamefont {Y.}~\bibnamefont {Zhou}}, \bibinfo {author}
  {\bibfnamefont {M.}~\bibnamefont {Ezawa}}, \bibinfo {author} {\bibfnamefont
  {X.}~\bibnamefont {Liu}}, \bibinfo {author} {\bibfnamefont {S.}~\bibnamefont
  {Finizio}}, \bibinfo {author} {\bibfnamefont {J.}~\bibnamefont {Raabe}},
  \bibinfo {author} {\bibfnamefont {N.~J.}\ \bibnamefont {Lee}}, \bibinfo
  {author} {\bibfnamefont {S.-I.}\ \bibnamefont {Kim}}, \bibinfo {author}
  {\bibfnamefont {S.-Y.}\ \bibnamefont {Park}}, \bibinfo {author}
  {\bibfnamefont {Y.}~\bibnamefont {Kim}}, \bibinfo {author} {\bibfnamefont
  {J.-Y.}\ \bibnamefont {Kim}}, \bibinfo {author} {\bibfnamefont
  {D.}~\bibnamefont {Lee}}, \bibinfo {author} {\bibfnamefont {O.}~\bibnamefont
  {Lee}}, \bibinfo {author} {\bibfnamefont {J.~W.}\ \bibnamefont {Choi}},
  \bibinfo {author} {\bibfnamefont {B.-C.}\ \bibnamefont {Min}}, \bibinfo
  {author} {\bibfnamefont {H.~C.}\ \bibnamefont {Koo}},\ and\ \bibinfo {author}
  {\bibfnamefont {J.}~\bibnamefont {Chang}},\ }\bibfield  {title} {\bibinfo
  {title} {Current-driven dynamics and inhibition of the skyrmion {{Hall}}
  effect of ferrimagnetic skyrmions in {{GdFeCo}} films},\ }\href
  {https://doi.org/10.1038/s41467-018-03378-7} {\bibfield  {journal} {\bibinfo
  {journal} {Nature Communications}\ }\textbf {\bibinfo {volume} {9}},\
  \bibinfo {pages} {959} (\bibinfo {year} {2018})}\BibitemShut {NoStop}%
\bibitem [{\citenamefont {Zhang}\ \emph {et~al.}(2016)\citenamefont {Zhang},
  \citenamefont {Zhou},\ and\ \citenamefont
  {Ezawa}}]{zhangAntiferromagneticSkyrmionStability2016}%
  \BibitemOpen
  \bibfield  {author} {\bibinfo {author} {\bibfnamefont {X.}~\bibnamefont
  {Zhang}}, \bibinfo {author} {\bibfnamefont {Y.}~\bibnamefont {Zhou}},\ and\
  \bibinfo {author} {\bibfnamefont {M.}~\bibnamefont {Ezawa}},\ }\bibfield
  {title} {\bibinfo {title} {Antiferromagnetic {{Skyrmion}}: {{Stability}},
  {{Creation}} and {{Manipulation}}},\ }\href
  {https://doi.org/10.1038/srep24795} {\bibfield  {journal} {\bibinfo
  {journal} {Scientific Reports}\ }\textbf {\bibinfo {volume} {6}},\ \bibinfo
  {pages} {24795} (\bibinfo {year} {2016})}\BibitemShut {NoStop}%
\bibitem [{\citenamefont {Jiang}\ \emph {et~al.}(2017)\citenamefont {Jiang},
  \citenamefont {Zhang}, \citenamefont {Yu}, \citenamefont {Zhang},
  \citenamefont {Wang}, \citenamefont {Benjamin~Jungfleisch}, \citenamefont
  {Pearson}, \citenamefont {Cheng}, \citenamefont {Heinonen}, \citenamefont
  {Wang}, \citenamefont {Zhou}, \citenamefont {Hoffmann},\ and\ \citenamefont
  {{te~Velthuis}}}]{jiangDirectObservationSkyrmion2017}%
  \BibitemOpen
  \bibfield  {author} {\bibinfo {author} {\bibfnamefont {W.}~\bibnamefont
  {Jiang}}, \bibinfo {author} {\bibfnamefont {X.}~\bibnamefont {Zhang}},
  \bibinfo {author} {\bibfnamefont {G.}~\bibnamefont {Yu}}, \bibinfo {author}
  {\bibfnamefont {W.}~\bibnamefont {Zhang}}, \bibinfo {author} {\bibfnamefont
  {X.}~\bibnamefont {Wang}}, \bibinfo {author} {\bibfnamefont {M.}~\bibnamefont
  {Benjamin~Jungfleisch}}, \bibinfo {author} {\bibfnamefont {J.~E.}\
  \bibnamefont {Pearson}}, \bibinfo {author} {\bibfnamefont {X.}~\bibnamefont
  {Cheng}}, \bibinfo {author} {\bibfnamefont {O.}~\bibnamefont {Heinonen}},
  \bibinfo {author} {\bibfnamefont {K.~L.}\ \bibnamefont {Wang}}, \bibinfo
  {author} {\bibfnamefont {Y.}~\bibnamefont {Zhou}}, \bibinfo {author}
  {\bibfnamefont {A.}~\bibnamefont {Hoffmann}},\ and\ \bibinfo {author}
  {\bibfnamefont {S.~G.~E.}\ \bibnamefont {{te~Velthuis}}},\ }\bibfield
  {title} {\bibinfo {title} {Direct observation of the skyrmion {{Hall}}
  effect},\ }\href {https://doi.org/10.1038/nphys3883} {\bibfield  {journal}
  {\bibinfo  {journal} {Nature Physics}\ }\textbf {\bibinfo {volume} {13}},\
  \bibinfo {pages} {162} (\bibinfo {year} {2017})}\BibitemShut {NoStop}%
\bibitem [{\citenamefont {Gnoli}\ \emph {et~al.}(2021)\citenamefont {Gnoli},
  \citenamefont {Riente}, \citenamefont {Vacca}, \citenamefont {Ruo~Roch},\
  and\ \citenamefont {Graziano}}]{gnoliSkyrmionLogicInMemoryArchitecture2021}%
  \BibitemOpen
  \bibfield  {author} {\bibinfo {author} {\bibfnamefont {L.}~\bibnamefont
  {Gnoli}}, \bibinfo {author} {\bibfnamefont {F.}~\bibnamefont {Riente}},
  \bibinfo {author} {\bibfnamefont {M.}~\bibnamefont {Vacca}}, \bibinfo
  {author} {\bibfnamefont {M.}~\bibnamefont {Ruo~Roch}},\ and\ \bibinfo
  {author} {\bibfnamefont {M.}~\bibnamefont {Graziano}},\ }\bibfield  {title}
  {\bibinfo {title} {Skyrmion {{Logic-In-Memory Architecture}} for
  {{Maximum}}/{{Minimum Search}}},\ }\href
  {https://doi.org/10.3390/electronics10020155} {\bibfield  {journal} {\bibinfo
   {journal} {Electronics}\ }\textbf {\bibinfo {volume} {10}},\ \bibinfo
  {pages} {155} (\bibinfo {year} {2021})}\BibitemShut {NoStop}%
\bibitem [{\citenamefont {Thiele}(1973)}]{thieleSteadyStateMotionMagnetic1973}%
  \BibitemOpen
  \bibfield  {author} {\bibinfo {author} {\bibfnamefont {A.~A.}\ \bibnamefont
  {Thiele}},\ }\bibfield  {title} {\bibinfo {title} {Steady-{{State Motion}} of
  {{Magnetic Domains}}},\ }\href {https://doi.org/10.1103/PhysRevLett.30.230}
  {\bibfield  {journal} {\bibinfo  {journal} {Physical Review Letters}\
  }\textbf {\bibinfo {volume} {30}},\ \bibinfo {pages} {230} (\bibinfo {year}
  {1973})}\BibitemShut {NoStop}%
\bibitem [{\citenamefont
  {Dzyaloshinsky}(1958)}]{dzyaloshinskyThermodynamicTheoryWeak1958}%
  \BibitemOpen
  \bibfield  {author} {\bibinfo {author} {\bibfnamefont {I.}~\bibnamefont
  {Dzyaloshinsky}},\ }\bibfield  {title} {\bibinfo {title} {A thermodynamic
  theory of ``weak'' ferromagnetism of antiferromagnetics},\ }\href
  {https://doi.org/10.1016/0022-3697(58)90076-3} {\bibfield  {journal}
  {\bibinfo  {journal} {Journal of Physics and Chemistry of Solids}\ }\textbf
  {\bibinfo {volume} {4}},\ \bibinfo {pages} {241} (\bibinfo {year}
  {1958})}\BibitemShut {NoStop}%
\bibitem [{\citenamefont
  {Moriya}(1960)}]{moriyaAnisotropicSuperexchangeInteraction1960}%
  \BibitemOpen
  \bibfield  {author} {\bibinfo {author} {\bibfnamefont {T.}~\bibnamefont
  {Moriya}},\ }\bibfield  {title} {\bibinfo {title} {Anisotropic
  {{Superexchange Interaction}} and {{Weak Ferromagnetism}}},\ }\href
  {https://doi.org/10.1103/PhysRev.120.91} {\bibfield  {journal} {\bibinfo
  {journal} {Physical Review}\ }\textbf {\bibinfo {volume} {120}},\ \bibinfo
  {pages} {91} (\bibinfo {year} {1960})}\BibitemShut {NoStop}%
\bibitem [{\citenamefont {Chen}\ \emph {et~al.}(2018)\citenamefont {Chen},
  \citenamefont {Kang}, \citenamefont {Zhu}, \citenamefont {Zhang},
  \citenamefont {Lei}, \citenamefont {Zhang}, \citenamefont {Zhou},\ and\
  \citenamefont {Zhao}}]{chenCompactSkyrmionicLeaky2018}%
  \BibitemOpen
  \bibfield  {author} {\bibinfo {author} {\bibfnamefont {X.}~\bibnamefont
  {Chen}}, \bibinfo {author} {\bibfnamefont {W.}~\bibnamefont {Kang}}, \bibinfo
  {author} {\bibfnamefont {D.}~\bibnamefont {Zhu}}, \bibinfo {author}
  {\bibfnamefont {X.}~\bibnamefont {Zhang}}, \bibinfo {author} {\bibfnamefont
  {N.}~\bibnamefont {Lei}}, \bibinfo {author} {\bibfnamefont {Y.}~\bibnamefont
  {Zhang}}, \bibinfo {author} {\bibfnamefont {Y.}~\bibnamefont {Zhou}},\ and\
  \bibinfo {author} {\bibfnamefont {W.}~\bibnamefont {Zhao}},\ }\bibfield
  {title} {\bibinfo {title} {A compact skyrmionic leaky\textendash
  integrate\textendash fire spiking neuron device},\ }\href
  {https://doi.org/10.1039/C7NR09722K} {\bibfield  {journal} {\bibinfo
  {journal} {Nanoscale}\ }\textbf {\bibinfo {volume} {10}},\ \bibinfo {pages}
  {6139} (\bibinfo {year} {2018})}\BibitemShut {NoStop}%
\bibitem [{\citenamefont {Zhang}\ and\ \citenamefont
  {Li}(2004)}]{zhangRolesNonequilibriumConduction2004}%
  \BibitemOpen
  \bibfield  {author} {\bibinfo {author} {\bibfnamefont {S.}~\bibnamefont
  {Zhang}}\ and\ \bibinfo {author} {\bibfnamefont {Z.}~\bibnamefont {Li}},\
  }\bibfield  {title} {\bibinfo {title} {Roles of {{Nonequilibrium Conduction
  Electrons}} on the {{Magnetization Dynamics}} of {{Ferromagnets}}},\ }\href
  {https://doi.org/10.1103/PhysRevLett.93.127204} {\bibfield  {journal}
  {\bibinfo  {journal} {Physical Review Letters}\ }\textbf {\bibinfo {volume}
  {93}},\ \bibinfo {pages} {127204} (\bibinfo {year} {2004})}\BibitemShut
  {NoStop}%
\bibitem [{\citenamefont
  {Gilbert}(2004)}]{gilbertphenomenologicaltheorydamping2004}%
  \BibitemOpen
  \bibfield  {author} {\bibinfo {author} {\bibfnamefont {T.}~\bibnamefont
  {Gilbert}},\ }\bibfield  {title} {\bibinfo {title} {A phenomenological theory
  of damping in ferromagnetic materials},\ }\href
  {https://doi.org/10.1109/TMAG.2004.836740} {\bibfield  {journal} {\bibinfo
  {journal} {IEEE Transactions on Magnetics}\ }\textbf {\bibinfo {volume}
  {40}},\ \bibinfo {pages} {3443} (\bibinfo {year} {2004})}\BibitemShut
  {NoStop}%
\bibitem [{\citenamefont {Thiaville}\ \emph {et~al.}(2005)\citenamefont
  {Thiaville}, \citenamefont {Nakatani}, \citenamefont {Miltat},\ and\
  \citenamefont
  {Suzuki}}]{thiavilleMicromagneticunderstandingcurrentdriven2005a}%
  \BibitemOpen
  \bibfield  {author} {\bibinfo {author} {\bibfnamefont {A.}~\bibnamefont
  {Thiaville}}, \bibinfo {author} {\bibfnamefont {Y.}~\bibnamefont {Nakatani}},
  \bibinfo {author} {\bibfnamefont {J.}~\bibnamefont {Miltat}},\ and\ \bibinfo
  {author} {\bibfnamefont {Y.}~\bibnamefont {Suzuki}},\ }\bibfield  {title}
  {\bibinfo {title} {Micromagnetic understanding of current-driven domain wall
  motion in patterned nanowires},\ }\href
  {https://doi.org/10.1209/epl/i2004-10452-6} {\bibfield  {journal} {\bibinfo
  {journal} {EPL (Europhysics Letters)}\ }\textbf {\bibinfo {volume} {69}},\
  \bibinfo {pages} {990} (\bibinfo {year} {2005})}\BibitemShut {NoStop}%
\bibitem [{\citenamefont {Everschor}\ \emph {et~al.}(2011)\citenamefont
  {Everschor}, \citenamefont {Garst}, \citenamefont {Duine},\ and\
  \citenamefont {Rosch}}]{everschorCurrentinducedRotationalTorques2011}%
  \BibitemOpen
  \bibfield  {author} {\bibinfo {author} {\bibfnamefont {K.}~\bibnamefont
  {Everschor}}, \bibinfo {author} {\bibfnamefont {M.}~\bibnamefont {Garst}},
  \bibinfo {author} {\bibfnamefont {R.~A.}\ \bibnamefont {Duine}},\ and\
  \bibinfo {author} {\bibfnamefont {A.}~\bibnamefont {Rosch}},\ }\bibfield
  {title} {\bibinfo {title} {Current-induced rotational torques in the skyrmion
  lattice phase of chiral magnets},\ }\href
  {https://doi.org/10.1103/PhysRevB.84.064401} {\bibfield  {journal} {\bibinfo
  {journal} {Physical Review B}\ }\textbf {\bibinfo {volume} {84}},\ \bibinfo
  {pages} {064401} (\bibinfo {year} {2011})}\BibitemShut {NoStop}%
\bibitem [{\citenamefont {El{\'i}as}\ \emph {et~al.}(2017)\citenamefont
  {El{\'i}as}, \citenamefont {{Vidal-Silva}},\ and\ \citenamefont
  {Manchon}}]{eliasSteadymotionskyrmions2017}%
  \BibitemOpen
  \bibfield  {author} {\bibinfo {author} {\bibfnamefont {R.~G.}\ \bibnamefont
  {El{\'i}as}}, \bibinfo {author} {\bibfnamefont {N.}~\bibnamefont
  {{Vidal-Silva}}},\ and\ \bibinfo {author} {\bibfnamefont {A.}~\bibnamefont
  {Manchon}},\ }\bibfield  {title} {\bibinfo {title} {Steady motion of
  skyrmions and domains walls under diffusive spin torques},\ }\href
  {https://doi.org/10.1103/PhysRevB.95.104406} {\bibfield  {journal} {\bibinfo
  {journal} {Physical Review B}\ }\textbf {\bibinfo {volume} {95}},\ \bibinfo
  {pages} {104406} (\bibinfo {year} {2017})}\BibitemShut {NoStop}%
\bibitem [{\citenamefont {Huber}(1982)}]{HuberDynamicsspinvortices1982}%
  \BibitemOpen
  \bibfield  {author} {\bibinfo {author} {\bibfnamefont {D.~L.}\ \bibnamefont
  {Huber}},\ }\bibfield  {title} {\bibinfo {title} {Dynamics of spin vortices
  in two-dimensional planar magnets},\ }\href
  {https://doi.org/10.1103/PhysRevB.26.3758} {\bibfield  {journal} {\bibinfo
  {journal} {Physical Review B}\ }\textbf {\bibinfo {volume} {26}},\ \bibinfo
  {pages} {3758} (\bibinfo {year} {1982})}\BibitemShut {NoStop}%
\bibitem [{\citenamefont {Iwasaki}\ \emph {et~al.}(2013)\citenamefont
  {Iwasaki}, \citenamefont {Mochizuki},\ and\ \citenamefont
  {Nagaosa}}]{iwasakiUniversalCurrentvelocityRelation2013}%
  \BibitemOpen
  \bibfield  {author} {\bibinfo {author} {\bibfnamefont {J.}~\bibnamefont
  {Iwasaki}}, \bibinfo {author} {\bibfnamefont {M.}~\bibnamefont {Mochizuki}},\
  and\ \bibinfo {author} {\bibfnamefont {N.}~\bibnamefont {Nagaosa}},\
  }\bibfield  {title} {\bibinfo {title} {Universal current-velocity relation of
  skyrmion motion in chiral magnets},\ }\href
  {https://doi.org/10.1038/ncomms2442} {\bibfield  {journal} {\bibinfo
  {journal} {Nature Communications}\ }\textbf {\bibinfo {volume} {4}},\
  \bibinfo {pages} {1463} (\bibinfo {year} {2013})}\BibitemShut {NoStop}%
\bibitem [{\citenamefont {Tomasello}\ \emph {et~al.}(2014)\citenamefont
  {Tomasello}, \citenamefont {Martinez}, \citenamefont {Zivieri}, \citenamefont
  {Torres}, \citenamefont {Carpentieri},\ and\ \citenamefont
  {Finocchio}}]{tomaselloStrategyDesignSkyrmion2014}%
  \BibitemOpen
  \bibfield  {author} {\bibinfo {author} {\bibfnamefont {R.}~\bibnamefont
  {Tomasello}}, \bibinfo {author} {\bibfnamefont {E.}~\bibnamefont {Martinez}},
  \bibinfo {author} {\bibfnamefont {R.}~\bibnamefont {Zivieri}}, \bibinfo
  {author} {\bibfnamefont {L.}~\bibnamefont {Torres}}, \bibinfo {author}
  {\bibfnamefont {M.}~\bibnamefont {Carpentieri}},\ and\ \bibinfo {author}
  {\bibfnamefont {G.}~\bibnamefont {Finocchio}},\ }\bibfield  {title} {\bibinfo
  {title} {A strategy for the design of skyrmion racetrack memories},\ }\href
  {https://doi.org/10.1038/srep06784} {\bibfield  {journal} {\bibinfo
  {journal} {Scientific Reports}\ }\textbf {\bibinfo {volume} {4}},\ \bibinfo
  {pages} {6784} (\bibinfo {year} {2014})}\BibitemShut {NoStop}%
\bibitem [{\citenamefont {Wang}\ \emph {et~al.}(2022)\citenamefont {Wang},
  \citenamefont {Song}, \citenamefont {Wei}, \citenamefont {Nan}, \citenamefont
  {Zhang}, \citenamefont {Ge}, \citenamefont {Tian}, \citenamefont {Zang},\
  and\ \citenamefont {Du}}]{wangElectricalManipulationSkyrmions2022}%
  \BibitemOpen
  \bibfield  {author} {\bibinfo {author} {\bibfnamefont {W.}~\bibnamefont
  {Wang}}, \bibinfo {author} {\bibfnamefont {D.}~\bibnamefont {Song}}, \bibinfo
  {author} {\bibfnamefont {W.}~\bibnamefont {Wei}}, \bibinfo {author}
  {\bibfnamefont {P.}~\bibnamefont {Nan}}, \bibinfo {author} {\bibfnamefont
  {S.}~\bibnamefont {Zhang}}, \bibinfo {author} {\bibfnamefont
  {B.}~\bibnamefont {Ge}}, \bibinfo {author} {\bibfnamefont {M.}~\bibnamefont
  {Tian}}, \bibinfo {author} {\bibfnamefont {J.}~\bibnamefont {Zang}},\ and\
  \bibinfo {author} {\bibfnamefont {H.}~\bibnamefont {Du}},\ }\bibfield
  {title} {\bibinfo {title} {Electrical manipulation of skyrmions in a chiral
  magnet},\ }\href {https://doi.org/10.1038/s41467-022-29217-4} {\bibfield
  {journal} {\bibinfo  {journal} {Nature Communications}\ }\textbf {\bibinfo
  {volume} {13}},\ \bibinfo {pages} {1593} (\bibinfo {year}
  {2022})}\BibitemShut {NoStop}%
\bibitem [{\citenamefont {Foster}\ \emph {et~al.}(2019)\citenamefont {Foster},
  \citenamefont {Kind}, \citenamefont {Ackerman}, \citenamefont {Tai},
  \citenamefont {Dennis},\ and\ \citenamefont
  {Smalyukh}}]{fosterTwodimensionalSkyrmionBags2019}%
  \BibitemOpen
  \bibfield  {author} {\bibinfo {author} {\bibfnamefont {D.}~\bibnamefont
  {Foster}}, \bibinfo {author} {\bibfnamefont {C.}~\bibnamefont {Kind}},
  \bibinfo {author} {\bibfnamefont {P.~J.}\ \bibnamefont {Ackerman}}, \bibinfo
  {author} {\bibfnamefont {J.-S.~B.}\ \bibnamefont {Tai}}, \bibinfo {author}
  {\bibfnamefont {M.~R.}\ \bibnamefont {Dennis}},\ and\ \bibinfo {author}
  {\bibfnamefont {I.~I.}\ \bibnamefont {Smalyukh}},\ }\bibfield  {title}
  {\bibinfo {title} {Two-dimensional skyrmion bags in liquid crystals and
  ferromagnets},\ }\href {https://doi.org/10.1038/s41567-019-0476-x} {\bibfield
   {journal} {\bibinfo  {journal} {Nature Physics}\ }\textbf {\bibinfo {volume}
  {15}},\ \bibinfo {pages} {655} (\bibinfo {year} {2019})}\BibitemShut
  {NoStop}%
\bibitem [{\citenamefont {Kind}\ and\ \citenamefont
  {Foster}(2021)}]{kindMagneticSkyrmionBinning2021}%
  \BibitemOpen
  \bibfield  {author} {\bibinfo {author} {\bibfnamefont {C.}~\bibnamefont
  {Kind}}\ and\ \bibinfo {author} {\bibfnamefont {D.}~\bibnamefont {Foster}},\
  }\bibfield  {title} {\bibinfo {title} {Magnetic skyrmion binning},\ }\href
  {https://doi.org/10.1103/PhysRevB.103.L100413} {\bibfield  {journal}
  {\bibinfo  {journal} {Physical Review B}\ }\textbf {\bibinfo {volume}
  {103}},\ \bibinfo {pages} {L100413} (\bibinfo {year} {2021})}\BibitemShut
  {NoStop}%
\bibitem [{\citenamefont {Vansteenkiste}\ \emph {et~al.}(2014)\citenamefont
  {Vansteenkiste}, \citenamefont {Leliaert}, \citenamefont {Dvornik},
  \citenamefont {Helsen}, \citenamefont {{Garcia-Sanchez}},\ and\ \citenamefont
  {Van~Waeyenberge}}]{vansteenkisteDesignVerificationMuMax32014}%
  \BibitemOpen
  \bibfield  {author} {\bibinfo {author} {\bibfnamefont {A.}~\bibnamefont
  {Vansteenkiste}}, \bibinfo {author} {\bibfnamefont {J.}~\bibnamefont
  {Leliaert}}, \bibinfo {author} {\bibfnamefont {M.}~\bibnamefont {Dvornik}},
  \bibinfo {author} {\bibfnamefont {M.}~\bibnamefont {Helsen}}, \bibinfo
  {author} {\bibfnamefont {F.}~\bibnamefont {{Garcia-Sanchez}}},\ and\ \bibinfo
  {author} {\bibfnamefont {B.}~\bibnamefont {Van~Waeyenberge}},\ }\bibfield
  {title} {\bibinfo {title} {The design and verification of {{MuMax3}}},\
  }\href {https://doi.org/10.1063/1.4899186} {\bibfield  {journal} {\bibinfo
  {journal} {AIP Advances}\ }\textbf {\bibinfo {volume} {4}},\ \bibinfo {pages}
  {107133} (\bibinfo {year} {2014})}\BibitemShut {NoStop}%
\bibitem [{\citenamefont {Guang}\ \emph {et~al.}(2023)\citenamefont {Guang},
  \citenamefont {Zhang}, \citenamefont {Zhang}, \citenamefont {Wang},
  \citenamefont {Zhao}, \citenamefont {Tomasello}, \citenamefont {Zhang},
  \citenamefont {He}, \citenamefont {Li}, \citenamefont {Liu}, \citenamefont
  {Feng}, \citenamefont {Wei}, \citenamefont {Carpentieri}, \citenamefont
  {Hou}, \citenamefont {Liu}, \citenamefont {Peng}, \citenamefont {Zeng},
  \citenamefont {Finocchio}, \citenamefont {Zhang}, \citenamefont {Coey},
  \citenamefont {Han},\ and\ \citenamefont
  {Yu}}]{guangElectricalDetectionMagnetic2023}%
  \BibitemOpen
  \bibfield  {author} {\bibinfo {author} {\bibfnamefont {Y.}~\bibnamefont
  {Guang}}, \bibinfo {author} {\bibfnamefont {L.}~\bibnamefont {Zhang}},
  \bibinfo {author} {\bibfnamefont {J.}~\bibnamefont {Zhang}}, \bibinfo
  {author} {\bibfnamefont {Y.}~\bibnamefont {Wang}}, \bibinfo {author}
  {\bibfnamefont {Y.}~\bibnamefont {Zhao}}, \bibinfo {author} {\bibfnamefont
  {R.}~\bibnamefont {Tomasello}}, \bibinfo {author} {\bibfnamefont
  {S.}~\bibnamefont {Zhang}}, \bibinfo {author} {\bibfnamefont
  {B.}~\bibnamefont {He}}, \bibinfo {author} {\bibfnamefont {J.}~\bibnamefont
  {Li}}, \bibinfo {author} {\bibfnamefont {Y.}~\bibnamefont {Liu}}, \bibinfo
  {author} {\bibfnamefont {J.}~\bibnamefont {Feng}}, \bibinfo {author}
  {\bibfnamefont {H.}~\bibnamefont {Wei}}, \bibinfo {author} {\bibfnamefont
  {M.}~\bibnamefont {Carpentieri}}, \bibinfo {author} {\bibfnamefont
  {Z.}~\bibnamefont {Hou}}, \bibinfo {author} {\bibfnamefont {J.}~\bibnamefont
  {Liu}}, \bibinfo {author} {\bibfnamefont {Y.}~\bibnamefont {Peng}}, \bibinfo
  {author} {\bibfnamefont {Z.}~\bibnamefont {Zeng}}, \bibinfo {author}
  {\bibfnamefont {G.}~\bibnamefont {Finocchio}}, \bibinfo {author}
  {\bibfnamefont {X.}~\bibnamefont {Zhang}}, \bibinfo {author} {\bibfnamefont
  {J.~M.~D.}\ \bibnamefont {Coey}}, \bibinfo {author} {\bibfnamefont
  {X.}~\bibnamefont {Han}},\ and\ \bibinfo {author} {\bibfnamefont
  {G.}~\bibnamefont {Yu}},\ }\bibfield  {title} {\bibinfo {title} {Electrical
  {{Detection}} of {{Magnetic Skyrmions}} in a {{Magnetic Tunnel Junction}}},\
  }\href {https://doi.org/10.1002/aelm.202200570} {\bibfield  {journal}
  {\bibinfo  {journal} {Advanced Electronic Materials}\ }\textbf {\bibinfo
  {volume} {9}},\ \bibinfo {pages} {2200570} (\bibinfo {year}
  {2023})}\BibitemShut {NoStop}%
\bibitem [{\citenamefont
  {Abbott}(1999)}]{abbottLapicqueIntroductionIntegrateandfire1999}%
  \BibitemOpen
  \bibfield  {author} {\bibinfo {author} {\bibfnamefont {L.~F.}\ \bibnamefont
  {Abbott}},\ }\bibfield  {title} {\bibinfo {title} {Lapicque's introduction of
  the integrate-and-fire model neuron (1907)},\ }\href
  {https://doi.org/10.1016/S0361-9230(99)00161-6} {\bibfield  {journal}
  {\bibinfo  {journal} {Brain Research Bulletin}\ }\textbf {\bibinfo {volume}
  {50}},\ \bibinfo {pages} {303} (\bibinfo {year} {1999})}\BibitemShut
  {NoStop}%
\bibitem [{\citenamefont {Hodgkin}\ and\ \citenamefont
  {Huxley}(1990)}]{hodgkinQuantitativeDescriptionMembrane1990}%
  \BibitemOpen
  \bibfield  {author} {\bibinfo {author} {\bibfnamefont {A.~L.}\ \bibnamefont
  {Hodgkin}}\ and\ \bibinfo {author} {\bibfnamefont {A.~F.}\ \bibnamefont
  {Huxley}},\ }\bibfield  {title} {\bibinfo {title} {A quantitative description
  of membrane current and its application to conduction and excitation in
  nerve},\ }\href {https://doi.org/10.1007/BF02459568} {\bibfield  {journal}
  {\bibinfo  {journal} {Bulletin of Mathematical Biology}\ }\textbf {\bibinfo
  {volume} {52}},\ \bibinfo {pages} {25} (\bibinfo {year} {1990})}\BibitemShut
  {NoStop}%
\bibitem [{\citenamefont {Kind}\ \emph {et~al.}(2020)\citenamefont {Kind},
  \citenamefont {Friedemann},\ and\ \citenamefont
  {Read}}]{kindExistenceStabilitySkyrmion2020}%
  \BibitemOpen
  \bibfield  {author} {\bibinfo {author} {\bibfnamefont {C.}~\bibnamefont
  {Kind}}, \bibinfo {author} {\bibfnamefont {S.}~\bibnamefont {Friedemann}},\
  and\ \bibinfo {author} {\bibfnamefont {D.}~\bibnamefont {Read}},\ }\bibfield
  {title} {\bibinfo {title} {Existence and stability of skyrmion bags in thin
  magnetic films},\ }\href {https://doi.org/10.1063/1.5127173} {\bibfield
  {journal} {\bibinfo  {journal} {Applied Physics Letters}\ }\textbf {\bibinfo
  {volume} {116}},\ \bibinfo {pages} {022413} (\bibinfo {year}
  {2020})}\BibitemShut {NoStop}%
\end{thebibliography}%

\end{document}